\documentclass{aa}
\usepackage{psfig}

\begin{document}
   \title{Microlensing planets in M22: free-floating or bound?} 
   \titlerunning{Microlensing planets in M22: free-floating or bound?}
   \author{R. de la Fuente Marcos
           \and
           C. de la Fuente Marcos}
   \authorrunning{R. de la Fuente Marcos \& C. de la Fuente Marcos}
   \offprints{R. de la Fuente Marcos
             }
   \institute{Universidad Complutense de Madrid, Ciudad Universitaria, 
              E-28040 Madrid, Spain}  
   \date{Received 9 July 2001 / Accepted XX Xxxxxx 2001}

   \abstract{We use detailed numerical simulations and theoretical
             estimates to show that, if confirmed, the unusually brief 
             microlensing events observed by Sahu et al. (2001) in the 
             field of the globular cluster M22 might be explained as a 
             result of microlensing by a population of clustered MACHOs, 
             a dark cluster or RAMBO, not associated with the globular 
             cluster. If real, this dark cluster would be located 
             between M22 and the Galactic bulge and could include at 
             least $10^6$ substellar members with a typical size of 
             1-3 pc. Bound planets in wide or/and eccentric orbits are 
             also able to reproduce the observed microlensing behaviour, 
             but only if multiplanet systems (including large 
             Kuiper-belt-like objects) are abundant, although, our 
             calculations argue against the latter scenario as the 
             ionization rate in M22 is very high. Dynamically ejected or 
             lone planets are, in principle, incompatible with the 
             observational findings as they either escape their parent 
             cluster in a relatively short time-scale after ejection or 
             segregate toward the outskirts of the cluster. We discuss 
             additional implications of the dark cluster scenario, 
             including the existence of a population of RAMBOs toward 
             the Galactic bulge.
   }

   \maketitle

   \section{Introduction}
      Gravitational microlensing occurs when the gravitational field
      of a massive object (the lens), located close to the observer's
      line-of-sight, bends the light from a distant object (the source) 
      to generate two or more unresolved images. For a point mass event,
      the two images are not identical, they have different areas and
      opposite parities. Although the gravitational microlensing effect 
      conserves surface brightness it alters the flux of the unresolved images 
      as the relative positions of the lens, the source, and the observer
      change. These variations in flux translate into a measurable 
      increase of the photometric magnitude (magnification). The effect 
      does not depend on the photometric color used in the observations. 
      Microlensing was proposed in 1986 by Paczy\'nski as a method to 
      detect compact baryonic dark matter in the halo of our Galaxy, but 
      before the first events were discovered, Mao and Paczy\'nski (1991) 
      had already noted that is might be possible to detect planetary 
      companions of the primary microlenses.
      \hfil\par
      Stars or sub-stellar objects in globular clusters can act either as 
      sources to detect MACHOS (Massive Astrophysical Compact Halo Objects) 
      located along the line-of-sight or as lenses for more distant background 
      stars. Under normal conditions, the probability of detecting microlensing
      events is very small but observing a globular cluster projected against 
      the star-rich Galactic bulge increases this probability by a significant 
      amount. Monitoring the bulge stars for variability may then help to detect
      dim or dark objects within the star cluster. Using this technique,
      Sahu et al. (2001) (hereafter SM22) have recently presented evidence 
      for the existence of free-floating planetary-mass objects in M22. They
      have detected six microlensing events that are completely unresolved
      in time. If these events indeed represent gravitational microlensing,
      the upper limit to the mass of the lens is less than 80 $M_{\oplus}$ 
      (Saturn mass is about 95). They interpret that these objects must be
      either free-floating, or at least several AUs from any stellar-mass
      objects, although they favour the free-floating hypothesis. If real, 
      the total contribution of these free-floating planets to the mass of 
      M22 is estimated by these authors to be of the order of 10\%.
      \hfil\par
      M22 (NGC6656) is an unusual star cluster with a diameter of about 18 pc 
      and ranking fourth in brightness among globulars, this heavily 
      reddened metal-poor globular cluster is 12 Gyr old (Davidge \& 
      Harris 1996). Its binary fraction is 0.01-0.03 depending on the 
      eccentricity as opposed to 0.12, the binary fraction for nearby, 
      solar-type stars having similar mass ratios and periods 
      (Cote et al. 1996). M22 has a binary ionization rate as high as 
      $\omega$ Centauri and very likely all its soft (long period) 
      binaries have been disrupted by stellar encounters.
      \hfil\par
      In this paper, we investigate how bound planets can induce single
      lens gravitational microlensing in star clusters. In $\S$ 2, we present
      relevant results from numerical simulations. Different scenarios
      able to explain the SM22 findings are introduced in $\S$ 3 as well as
      relevant microlensing theory. $\S$ 4 is a discussion.

   \section{Models and results} 
      Here, we present partial results from a systematic long-term
      project aimed to study planetary dynamics in star clusters. Full details
      and results of our calculations will be presented elsewhere. Partial
      results from this program can be found in de la Fuente Marcos and de la 
      Fuente Marcos (1997, 1998, 2000, 2001a, b). In this paper, we 
      will only provide the data required to support our interpretation of the 
      results from SM22. Calculations have been carried out using a version of 
      Sverre Aarseth's {\small NBODY5} code (Aarseth 1994). This code includes 
      the effect of the Galactic tidal field and the mass loss due to stellar 
      evolution (Eggleton et al. 1989). Most of the models consider a realistic
      mass spectrum (Scalo 1986) in the range [0.08, 15.0] $M_{\odot}$. 
      Spherical symmetry and constant star density or Plummer models are 
      assumed for generating initial positions, with the ratio of the total 
      kinetic and potential energy fixed to 0.25 (0.5 for virial equilibrium). 
      Our choice of the initial ratio of kinetic to potential energy produces 
      an initial contraction of the cluster and simulates violent relaxation.
      Several models ($N$ = 1000 particles) include both 
      a significant primordial binary fraction and realistic orbital elements 
      for this binary population. The $N$ in our models ranges from 100 to 
      10,000. Although globular clusters are significantly more populated, our 
      present results are consistent with very preliminary partial results from
      our models with $N \approx 50,000$ and also with results from other 
      authors (Laughlin \& Adams 1998, 2000; Adams \& Laughlin 2001; 
      Smith \& Bonnell 2001; Bonnell et al. 2001), therefore we will apply them
      to globular clusters through this paper. Instead of computing the 
      evolution of several cluster models with different populations and sizes 
      we consider samples of models (10 or more for each $N$) only differing 
      in the value of the seed 
      number for generating initial conditions. For example, in one of our 
      samples we choose an open cluster with $N$ = 1000 objects (40\% binaries,
      60\% planetary systems) and a half-mass radius of about 0.9 pc. We select
      40\% primordial binaries as a plausible binary percentage for typical 
      galactic clusters, with semi-major axis in the range 51.6-309.4 AU. For 
      simplicity, the planetary systems studied in this research consist of 
      one giant planet and its host star. Our giant planet populations have 
      semi-major axes in the range [0.5, 60] AU, masses uniformly distributed 
      in the range 0.1-8 $M_{J}$ (9.6 $\times 10^{-4} \ M_{\odot}$), and an 
      initial eccentricity of 0.010. Pericentre, nodes, and inclinations for 
      both, planetary systems and binaries, as well as the eccentricities of 
      the binaries, are chosen from a random (thermalized) distribution. 
      \hfil\par
      Our calculations show that a significant fraction of the planetary 
      systems suffer modification of their primordial orbital elements as a 
      result of complex gravitational interactions with binaries, single stars,
      and/or other planetary systems. Eccentricity modification is, by far, 
      the most common event. Modification of the semi-major axis is not so
      frequent and just in a few cases an interaction results in a decrease of 
      the orbital size. Most of the times, the semi-major axis changes as a 
      result not only of external perturbations but also because of mass loss 
      from the parent star. The more massive the planet, the higher the 
      probability of being involved in an orbital modification event. 

   \section{Free-floating or bound?}
      Before presenting different scenarios able to induce planetary
      microlensing we introduce some relevant equations from the gravitational
      microlensing theory. Following Gaudi and Sackett (2000), let us consider 
      a single lens, in this case the time-variable flux observed from a 
      microlensed star is $F(t) = F_o (A(t) + f_B)$, where $F_o$ is the 
      unlensed flux of the star, $f_B$ is the ratio of any unresolved, unlensed
      background light to $F_o$ (the "blend fraction"), and $A(t)$ is the 
      magnification. The magnification for a single lens is given by 
      $A_o = (2 + u^2) / (u \sqrt{4 + u^2})$, 
      where $u$ is the instantaneous angular separation of the source and 
      the lens in units of the angular Einstein ring radius $\theta_E$
      of the lens that can be written as
      \begin{equation}
         \theta_E = \sqrt{\frac{4 G M}{c^2} \frac{D_{LS}}{D_{OL} D_{OS}}} \,,
          \label{einstein}
      \end{equation}
      where $G$ is the gravitational constant, $M$ is the mass of the lens,
      $c$ is the speed of light, and $D_{LS}$, $D_{OS}$, $D_{OL}$ are the
      lens-source, observer-source, and observer-lens distances, respectively.
      For a lens in M22 and a source in the Galactic bulge, $D_{OL}$ = 2.6 kpc
      and $D_{OS}$ = 8.2 kpc, therefore
      \begin{equation}
         \theta_E \approx 653 \ {\mu}as \ \ \sqrt{\frac{M}{0.2 M_{\odot}}} \,.
          \label{eM22}
      \end{equation} 
      In principle, if $u > 1.5$, the magnification is not large enough to 
      be measurable, although it depends strongly on the photometric errors.
      The characteristic timescale of one of these events is given by the
      Einstein time, $t_{E} = \theta_E \ D_{OL} / v_{T}$, where $v_T$ is the 
      transverse velocity of the lens relative to the observer-source 
      line-of-sight. This timescale increases as $\sqrt{M}$, 
      therefore the smaller the
      lens the shorter the duration of the microlensing event. If the lens is 
      not a single object but it has a companion, the formalism for binary 
      lenses should be used. The flux is still expressed by the same equation, 
      but the magnification can no longer be calculated analytically and 
      numerical techniques are required (Witt 1990). 
      The minimum separation between a planet and its host star (in units
      of $\theta_E$) for which planets can be discovered using microlensing
      is about 0.8. For planetary separations roughly between 0.8 and 1.5, 
      the microlensing signature corresponds to a binary lens. For larger
      values of the separation, the planet generally acts as an independent
      lens. The angular separation given by Eq. (\ref{eM22}) is equivalent 
      to the spatial distance
      \begin{equation}
         d_E \approx 1.70 \ AU \ \ \sqrt{\frac{M}{0.2 M_{\odot}}} \,.
         \label{distance}
      \end{equation} 
      In order to obtain isolated short-duration events corresponding to 
      planetary-class single-lens microlensing, the instantaneous separation 
      between the planet and its host star must be larger than 1.5 $d_E$. 
      This magnitude is plotted in Fig. \ref{esep}, therefore virtually 
      any planetary system (in M22) with a separation between planet and 
      host star wider than about 7 AU (upper limit for massive primary) will 
      induce a single lens microlensing event in exactly the same way a 
      lone planet does. As an example and from the figure, if the primary 
      has a mass of 0.3 $M_{\odot}$, the minimum distance for single lens 
      is 3.1 AU at the 95\% confidence level. For this hypothetical system any 
      planet farther from the central star may induce single lens microlensing.
      More rigorous arguments using the full binary lens formalism give
      essentially the same results. Let us consider several configurations 
      able to produce planetary-class single lens microlensing: 
%
%
     \begin{figure}[htbp]
        \psfig{figure=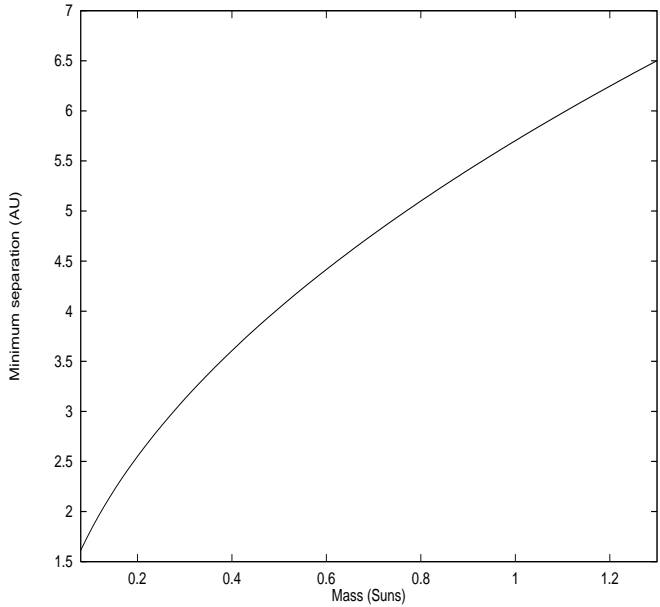,height=8cm,width=9cm,angle=-90}
        \caption[]{Einstein ring radius of the lens multiplied by
                   1.5 as a function of the stellar mass for M22. 
                   A bound planet with separation from its host 
                   star below the curve would induce double lens
                   behaviour with contributions of both objects, planet
                   and host star, clearly visible on the light curve.
                   Above the curve, single lens behaviour is expected.}
        \label{esep}
     \end{figure}
%

      \subsection{Free-floating planets: loners or escapees?}

       \subsubsection{Lone planets}
          SM22 interpret that their unresolved microlensing events have 
          been probably induced by a free-floating planetary-mass population
          located at the cluster core. However, existence of standalone 
          planets poses some theoretical challenges. Star clusters tend to 
          energy equipartition among stars of different mass. As a result of 
          this process, the heavier stars will tend to slow down and sink 
          toward the center of the cluster. Globular clusters are very old, 
          therefore mass segregation has already been well established 
          through the entire cluster. As an example from our calculations, for 
          a $N = 10,000$ model and after about 500 Myr mass segregation gives 
          an average mass of about 1 $M_{\odot}$ for the cluster core and 
          about 0.5 $M_{\odot}$ for the halo, and these values remain rather 
          constant during the evolution of the model. 
          Therefore, if they form, low-mass substellar objects 
          should be a minority in the cluster core. On the other hand, 
          standalone planets cannot be formed by accretion, therefore 
          protostellar collapse and fragmentation is currently the 
          only other alternative. Boss (2001) has shown that if the magnetic 
          field tension effects are important, collapse and fragmentation of 
          molecular clouds might be able to produce self-gravitating objects 
          with Jovian (not Saturn or smaller) masses although this author 
          terms the resulting objects as sub-brown dwarfs not planets. 
          However, it seems that, regardless of whether the planets were
          formed via collapse or were formed via accretion and subsequently
          ionized, they would still not survive in the core of the cluster.
          Free-floating planets have recently been identified in Orion
          (Lucas \& Roche 2000; Zapatero Osorio et al. 2000) but their
          origin remains controversial as they can be escapees or even actual
          brown dwarfs.
 
       \subsubsection{Wandering planets}
          Dynamic or supernova driven ejection of planets is possible
          in star clusters (de la Fuente Marcos \& de la Fuente Marcos 
          1998) but the fraction of runaway planets is less than a few
          percent for the entire cluster, much smaller than the figure 
          suggested by SM22. However, if the analysis is restricted to 
          the core of a globular cluster which is a very high star density
          region this number rises to about 50\%, although ejected planets 
          have very high characteristic velocities with the smaller planets 
          escaping faster. The actual percentage depends on the semi-major
          axis: if it is $< 3$ AU the ejection rate is 1\% but it is almost
          100\% if the semiaxis is $> 15$ AU. On the other hand, in our 
          models more than 70\% of planetary escapees have a velocity 
          $<$ 25 km/s with an upper limit for the remaining fraction
          of about 200 km/s. Fast or super-fast planetary escapees are always
          the result of strong multibody interactions inside the cluster core.
          Although for the core of M22 the escape velocity is in the range
          26-31 km/s, ejections from the core are the result of strong close
          encounters with characteristic velocities much higher than the
          escape velocity. Therefore, it is very unlikely to be able to 
          detect several of these events just by chance unless they are
          ejected planets moving almost parallel to the line-of-sight. In 
          that case, the number of detected events might be consistent with the 
          expected fraction of runaways from the cluster core. In this 
          scenario, the detected objects would be moving toward us and likely 
          located between M22 and the Earth as they were ejected long ago. 
          One can argue that this is practically impossible if the 
          ejection is isotropic, as the fraction of planets moving sufficiently
          close to our line-of-sight would be incredibly tiny; however, 
          ejections are only isotropical if they are the result of close 
          encounters (likely at the cluster core). If planetary objects escape 
          from the cluster as a result of evaporation (gradual increase in 
          kinetic energy due to distant encounters) then they escape through 
          the Lagrangian points, L$_1$ and L$_2$. If we consider lone planets 
          formed by collapse, this is a plausible scenario only if a Lagrangian
          point is along our line-of-sight. However and although the lack of 
          precise three-dimensional information makes it difficult to 
          estimate, it is not very likely that one of the Lagrangian points 
          might be projected toward the cluster core as the cluster is not 
          observed against the Galactic Center. 
          
      \subsection{Solar-System-like systems}
          Low eccentricity, distant sub-Saturnian planets similar to the
          gas giants (Uranus and Neptune) in our own Solar System are also 
          able to generate microlensing events like the ones described in SM22.
          However, the probability of getting them involved in microlensing 
          events is only significant if these planets are abundant (several 
          per host star). On the other hand, Edgeworth-Kuiper-belt-like 
          structures including Earth-sized objects in large numbers may also 
          be able to induce brief microlensing events. In fact and from a
          strictly intuitive point of view, the relatively high number of 
          planetary-class detections (6) as compared to a single 
          {\em classical} (stellar) event argues in favor of a multiplanet 
          scenario. However, low eccentricity primordial planetary orbits 
          are very unlikely in a highly ionizing environment like M22, at 
          least in the cluster core. As pointed out before, planetary systems
          wider than 10-15 AU are catastrophically disrupted and due to mass
          segregation the most massive stars, and likely less favorable for
          planetary-class microlensing (Fig. \ref{esep}), are dominant
          in the core and therefore the optical depth for bound planetary 
          microlensing at the cluster core is not very significant. 
          \hfil\par
          On the other hand, it is also possible that the detected objects are 
          projected against the core but they are actually part of the halo 
          of M22 where the star density is low enough to allow for relatively 
          primordial, unperturbed (likely multiplanet) systems. Our results 
          suggest that the percentage of planetary systems disrupted in the 
          halo of a typical star cluster is very negligible, and the fraction 
          that experiences significant variation ($>$ 10\%) in the orbital 
          elements is a few percent and mainly connected with stellar evolution
          not dynamical interactions. If that is the case, the number of 
          multiplanet systems similar to our Solar System in M22, and likely 
          in other globulars, could be much higher than expected. It is 
          possible to argue against this latter scenario claiming that the 
          optical depth (or probability) to lensing for the halo of M22 is 
          negligible. However, this could not be the case if we consider
          planetary objects instead of stars. Let us assume that the membership
          of M22 is $10^6$, simulations suggest that for rich star clusters the 
          core includes about 10\% of the cluster total population, therefore 
          we have about $10^5$ core stars (or planets if we assume that they 
          exist) able to contribute to microlensing. As pointed out before, 
          the actual number of planets available for single lens microlensing 
          could be just a small fraction of that number, as low as a few 
          percent.  However, we are observing through the cluster halo. If we 
          assume that the volume of the cluster is about $10^3$ the core volume
          and the core population is 10\% of the total population, then the 
          cluster average star density is about 1/100 the core density. On the 
          other hand, if we consider that the HST observed a region of radius 
          1 pc (core radius) and that the radius of M22 is about 18 pc, then 
          the number of halo stars included in an imaginary cylinder of radius 
          1 pc and length 36 pc could be estimated by considering that the 
          average mass ratio between core and cluster is 1.5. This number is 
          about 30\% the core population or 30,000 stars, most of them 
          low-mass. If we assume an average of 4 giant planets per star we 
          found about 120,000 objects available for microlensing. It implies 
          that the optical depth to planetary companions in the halo of M22 
          relative to its core is at least a factor 100 higher. Therefore, our 
          simplified calculation suggests that, as regards single lens planetary
          microlensing, the halo of M22 is dominant. 
          \hfil\par
          The probability $P$ or optical depth for a microlensing event 
          is given by $P = \pi \theta_E \sigma$, where $\sigma$ is the 
          surface density of objects. Therefore, the probability for a 
          core-induced {\em classical} (stellar) microlensing
          event is $P_c \propto M_c \sigma_c$, where $M_c$ is the average
          mass for stars in the core, and $\sigma_c$ is the surface density
          of stars toward the cluster core. In addition, the probability
          for a halo-induced planetary microlensing event will be 
          $P_p \propto M_p \sigma_p$, with $M_p$ the average mass for planets
          in the halo and $\sigma_p$ the surface density of planets toward
          the cluster halo; this can be calculated by using $\sigma_p = 
          N_p \sigma_h$, where $N_p$ is the number of planets per halo star 
          and $\sigma_h$ is the surface density of stars toward the cluster 
          halo. To estimate the actual number of planets per star we may use 
          $\sigma_c/\sigma_h \sim 3$ and $M_c/M_p \sim 10^3$. On the other 
          hand, if as suggested by the observations the proportion of available 
          planets/stars for microlensing is 6 to 1 then the average number of 
          planetary-size companions per halo star could be as high as 
          20,000 if we consider sub-Saturnian objects or about 2,000,000 if
          the lenses are Earth-sized. In any case, the results of this 
          quantitative analysis makes it difficult to attribute the observed 
          events to planetary objects in M22 unless planets in globular 
          clusters (or the early Universe for that matter) formed in a 
          fundamentally different manner than planets in the Galactic disk. 
          
      \subsection{Eccentric planets}
          If the objects detected by SM22 are actually located in the core
          of M22, low-eccentricity primordial planetary orbits are very
          unlikely if the initial semi-major axis is larger than about 0.5 AU,
          as they should be, otherwise a double lens microlensing event must
          be recorded, Fig. \ref{esep}. Our numerical results show that 
          within the cluster lifetime and for high star density environments 
          (10$^3$-10$^4$ $M_{\odot}$/pc$^{-3}$) like the core of M22 about
          50\% of systems are disrupted with the remaining fraction being
          characterized by high eccentricities ($e > 0.3$). A large fraction 
          of planets in very elongated orbits increases the probability of 
          single lens microlensing but not very significantly due to random
          orientations.  

      \subsection{Dark clusters: MACHOs and RAMBOs}
          Finally, the events reported by SM22 may have been induced
          by non-M22-related substellar objects located along the 
          line-of-sight. Although current Massive Astrophysical Compact 
          Halo Objects (MACHOs) observational results toward the
          Large Magellanic Cloud (LMC) exclude brown dwarfs as the
          primary constituent of the halo (Alcock et al. 1998), any
          hypothetical population of free-floating halo (single or 
          binary) brown dwarfs should have been born elsewhere, namely 
          in massive dark clusters or Robust Associations of Massive 
          Baryonic Objects (RAMBOs) (Moore \& Silk 1995). The dynamics 
          of these objects must be quite different from that of typical 
          star clusters. With a very narrow mass range, the evaporation 
          of these RAMBOs should be very slow as predicted by the 
          evolution of mono-component cluster models (de la Fuente Marcos 
          1995). These very long-lived objects may exist in large numbers. 
          On the other hand, the microlensing rate is $\Gamma \propto 
          \sqrt{M} \ N$, where $M$ is the mass of the lens and $N$ the 
          number of available lensing objects. SM22 found a single stellar 
          microlensing event (very likely M22-related) and six unresolved 
          events. If they are indeed due to substellar objects in the field 
          they must be clustered and a naive estimate suggests that the 
          dark cluster population must be at least 100 times larger than the 
          core population of M22, with the same apparent diameter. It means 
          that the dark cluster might include $10^7-10^8$ members with a 
          total mass of $10^4-10^5 \ M_{\odot}$. In principle, the object 
          could be located in front of or behind M22. For the same duration 
          (upper limit, 0.8 days), the lens is more massive if the dark 
          cluster is near the Galactic bulge and if its transverse velocity 
          relative to the observer-source line-of-sight is higher. In any case 
          the objects are well below 13 Jovian masses. The existence of a 
          clustered thick disk-like component of dark matter in the Milky Way 
          has been suggested by Sanchez-Salcedo (1997, 1999) and Kerins (1997).
 
   \section{Discussion}
      Although one can argue that the most likely explanation of the
      brief events toward M22 is simply that they are not due to 
      microlensing but, for example, to stellar variability, we will not
      include this conclusion in our discussion. Rather than consider
      astrophysical explanations as stellar variability, we will focus
      our discussion, assuming that the events are due to microlensing,
      on the astrodynamical (or kinematical) explanations. 
      Our analysis suggests that only two scenarios are able to explain
      the unresolved microlensing events observed toward M22: a cluster
      halo very rich in multiplanet systems (but Solar System-like) or
      a chance alignement with a dark cluster or RAMBO. The upper limit 
      for the mass of the substellar lenses found by SM22 is about 80 
      times that of Earth. In the first scenario, as microlensing detection 
      is totally by chance it means that the distribution of masses of 
      the planetary-mass population in M22 peaks around that value with a 
      very likely negligible fraction of planets above it. However, in the 
      Galactic disk the fraction of substellar objects with Jovian (or 
      higher) masses is not negligible. If the objects are indeed bound, 
      most planetary systems in M22 are dominated by planets of about 
      Saturn's mass. Preferential formation of low-mass giant planets 
      in globular clusters can be explained as a result of poor metallicity 
      and shorter lifetime of protoplanetary disks. In any case, planetary 
      formation in the early Universe seems to be rather different from 
      current one with super-Jupiters being more numerous now. On the other
      hand, as SM22 detected only one {\em classical} (stellar) microlensing 
      event but six unexpectedly brief events it means that M22 is rather 
      rich in planetary-mass objects. As pointed out before, a planet in
      a low-eccentricity, short semi-major axis orbit will induce a
      double lens behaviour with both objects (planet \& host star) 
      contributing to the magnification. On the other hand, a planet
      in a wide but not eccentric orbit will be able to induce a photometric
      behaviour like the one observed by SM22 but these objects are unlikely
      in the core of a globular cluster. If the lenses are located at the
      cluster halo the relatively high rate of detections suggests that
      multiplanet systems are very common in globular clusters and that
      giant planets in wide orbits are dominant as in our own planetary system.
      \hfil\par
      The dark cluster scenario is also able to explain the observations.
      The only problem posed by this explanation is the mass of the
      cluster members. From the Einstein time equation, the upper limit
      for the mass of the lenses is a few Jovian masses. From a strictly
      theoretical point of view it is difficult to explain how such an
      object can form. On the other hand, the mass of the dark cluster
      might span the range $10^4-10^5 \ M_{\odot}$. This is consistent with
      primordial (pre-Galactic) origin of MACHO clusters as cosmological 
      considerations of the minimum Jeans mass suggest a typical pre-Galactic 
      mass scale of $10^4-10^6 \ M_{\odot}$. If there exist many more
      of these dark clusters, the microlensing statistics would be
      essentially unchanged from the unclustered case (Metcalf \& Silk 1996),
      therefore it would be a really serendipitous discovery. This is 
      consistent with the lack of an analogous population of events toward 
      the LMC (Alcock et al. 1998). 
      \hfil\par
      In this paper we have presented several plausible alternatives to the 
      {\em free-floating planets} interpretation of the microlensing events
      observed by SM22. Our analysis is not meant to provide a unique or
      complete model either to explain microlensing in M22 or in any other
      star cluster. Instead we simply point out different but compatible 
      scenarios able to generate the same photometric signatures as detected 
      by SM22. Our analysis can be easily tested by surveying low ionization 
      rate globular clusters like M71, M4, or NGC3201 in which, if they form, 
      a fraction of low eccentricity, short semi-major axis planetary systems 
      is likely to survive and be able to contribute to (binary) microlensing. 
      If the observation of other clusters is not successful in finding
      microlensing events then the dark cluster scenario is the only plausible
      scenario. In addition, a comparison between microlensing data from core 
      and off-core observations of M22 may also help to clarify this matter, 
      and follow-up HST observations can shed new light on this tentative 
      discovery.

   \begin{acknowledgements}
      We thank Dr. S. J. Aarseth for providing his computer code and
      Dr. K. C. Sahu for some remarks on his results. The authors thank the 
      Department of Astrophysics of Universidad Complutense de Madrid 
      (UCM) for allotting excellent computing facilities. We also thank
      the referee, Scott Gaudi, for his rapid and very helpful report. 
      Part of the calculations were performed on the SGI Origin 2000 of the 
      'Centro de Supercomputaci\'on Complutense' through the UCM project 
      'Din\'amica Estelar y Sistemas Planetarios' (CIP 454). In preparation 
      of this paper, we made use of the ASTRO--PH e-print server and the NASA 
      Astrophysics Data System. 
   \end{acknowledgements}


\begin{thebibliography}{}
    \bibitem{} Aarseth, S. J. 1994,
               in Galactic Dynamics and N-body Simulations,
               eds.\ G. Contopoulos, N. K. Spyrou, \& L. Vlahos,
               Springer Verlag, Berlin, p.\ 277 
    \bibitem{} Adams, F. C., \& Laughlin, G. 2001,
               Icarus, 150, 151
    \bibitem{} Alcock, C., Allsman, R. A., Alves, D., et al. 1998,
               ApJ, 499, L9
    \bibitem{} Bonnell, I. A., Smith, K. W., Davis, M. B., \& Horne, K. 2001,
               MNRAS, 322, 859
    \bibitem{} Boss, A. P. 2001,
               ApJ, 551, L167
    \bibitem{} Cote, P., Pryor, C., McClure, R. D., Fletcher, J. M., \&
               Hesser, J. E. 1996,
               AJ, 112, 574
    \bibitem{} Davidge, T. J., \& Harris, W. E. 1996,
               ApJ, 462, 255
    \bibitem{} Eggleton, P. P., Fitchett, M. J., \& Tout C. A. 1989,
               ApJ, 347, 998 
    \bibitem{} de la Fuente Marcos, R. 1995,
               A\&A, 301, 407
    \bibitem{} de la Fuente Marcos, C., \& de la Fuente Marcos, R. 1997, 
               A\&A, 326, L21
    \bibitem{} de la Fuente Marcos, C., \& de la Fuente Marcos, R. 1998, 
               NewA, 4, 21
    \bibitem{} de la Fuente Marcos, R., \& de la Fuente Marcos, C. 2000, 
               in Stellar Clusters and Associations: Convection, Rotation, and
               Dynamos, eds.\ R. Pallavicini, G. Micela, \& S. Sciortino, 
               ASP Conference Series, San Francisco, vol.\ 198, p.\ 183 
    \bibitem{} de la Fuente Marcos, C., \& de la Fuente Marcos, R. 2001a, 
               A\&A, 371, 1097
    \bibitem{} de la Fuente Marcos, C., \& de la Fuente Marcos, R. 2001b, 
               in Modes of Star Formation and the Origin of Field Populations,
               eds.\ E. Grebel, \& W. Brandner, 
               ASP Conference Series, San Francisco, in press
    \bibitem{} Gaudi, D. S., \& Sackett, P. D. 2000,
               ApJ, 528, 56
    \bibitem{} Kerins, E. J. 1997,
               A\&A, 322, 709
    \bibitem{} Laughlin, G., \& Adams, F. C. 1998,
               ApJ, 508, L171
    \bibitem{} Laughlin, G., \& Adams, F. C. 2000,
               Icarus, 145, 614
    \bibitem{} Lucas, P. W., \& Roche, P. F. 2000,
               MNRAS, 314, 858
    \bibitem{} Mao, S., \& Paczy\'nski, B. 1991,
               ApJ, 374, 37
    \bibitem{} Metcalf, R. B., \& Silk, J. 1996,
               ApJ 464, 218
    \bibitem{} Moore, B., \& Silk, J. 1995, 
               ApJ, 442, L5
    \bibitem{} Paczy\'nski, B. 1986,
               ApJ, 304, 1
    \bibitem{} Sahu, K. C., Casertano, S., Livio, M., et al. (SM22) 2001,
               Nat, 411, 1022
    \bibitem{} S\'anchez-Salcedo, F. J. 1997,
               ApJ, 487, L61
    \bibitem{} S\'anchez-Salcedo, F. J. 1999,
               MNRAS, 303, 755
    \bibitem{} Scalo, M. J. 1986,
               Fundam. Cosmic. Phys., 11, 1
    \bibitem{} Smith, K. W., \& Bonnell, I. A. 2001,
               MNRAS, 322, L1
    \bibitem{} Witt, H. J. 1990,
               A\&A, 236, 311
    \bibitem{} Zapatero Osorio, M. R., B\'ejar, V, J. S., Mart\'{\i}n, E. L.,
               et al. 2000,
               Sci, 290, 103
   \end{thebibliography}
\end{document}